\documentstyle[preprint,aps]{revtex}
\begin{document}

\draft

\title{Low Temperature Behavior of the Kondo Effect in Tomonaga-Luttinger
Liquid }

\author{Yu-Liang Liu}
\address{Max-Planck-Institut f\"{u}r Physik Komplexer Systeme, Bayreuther
Str. 40, D-01187 Dresden, Germany.}

\maketitle

\begin{abstract}

Using the bosonization method, we study the low temperature
behavior of the Kondo effect in the Tomonaga-Luttinger liquid and clearly
show that the power law temperature dependence of the impurity susceptibility
is completely determined by the repulsive electron-electron interaction
existing in the total spin channel and is independent of the electron-electron
interaction existing in the charge channels.

\end{abstract}
\vspace{1cm}


\newpage

Recently, 
the quantum impurity scattering of the Tomonaga-Luttinger(TL) liquid has been
extensively studied by using different techniques[\cite{1}-\cite{16}]. 
However, there is some controversy on the treatment of backward scattering of 
the conduction electrons on a quantum impurity or impurity-like hole in 
the valence band, although we all agree that the backward scattering 
drastically changes the behavior of a TL-liquid. 
The main difficulty is that we have not a reasonable non-perturbation
technique to treat strongly correlated systems such as
in the high energy physics and condensed matter physics. Another interesting
problem is a quantum magnetic impurity scattering of the TL-liquid (Kondo
model in one-dimensional electronic system). Although it is extensively
studied by many authors[\cite{19}-\cite{22}], the low energy and low
temperature properties of the Kondo effect in the TL-liquid still remain an
open problem because of the strong coupling between the impurity spin and the
conduction electrons in the low energy and low temperature limit.

In this paper, using the bosonization method, we give a detail study on the low
temperature behavior of the Kondo effect in the TL-liquid and first time
clearly show that the power law non-Fermi liquid behavior of the impurity
susceptibility completely depends upon the electron-electron 
interaction existing
in the total spin channel and is independent of the electron-electron
interaction existing in the charge channels. For a free electron system, 
$g=1$ ($g$ is a dimensionless coupling strength
parameter, $g=1$ corresponding to the free electron gas), the impurity
part shows a  Fermi liquid behavior. For a weak electron-electron interaction,
$g_{c}<g<1$, $g_{c}$ is
defined as satisfying relation: $(1-g^{2}_{c})^{2}=2g_{c}$, the impurity
susceptibility has a power law temperature dependence. For a strong repulsive
electron-electron interaction, $g\leq g_{c}$, the impurity susceptibility
satisfies the Curie law and the impurity fermion has a free fermion Green
function. Therefore, for the repulsive electron-electron interaction case, the
impurity susceptibility shows a non-Fermi liquid behavior. This surprising
behavior of the impurity spin in the TL-liquid is completely determined by the
strongly coupling fixed point Hamiltonian. However, in the strong coupling 
limit, we may have either the Fermi liquid
fixed point or the non-Fermi liquid fixed point which completely depends upon
the electron-electron interaction existing in the total spin channel. For an
one-dimensional free electron system, $g=1$, the system reduces into an
anisotropic two-channel Kondo model, the backward scattering potential
provides the channel anisotropy which completely destroys the non-Fermi liquid
behavior of the isotropic two-channel Kondo model and makes the system
have the Fermi liquid behavior.

We choose the following Hamiltonian to describe the Kondo effect in an 
one-dimensional interacting electronic system 
\begin{equation}
H_{T}=H_{0}+H_{I}+H_{K}
\end{equation}
\begin{equation}
H_{0}=-iv_{F}\int dx(\psi^{+}_{R\sigma}(x)\partial_{x}\psi_{R\sigma}(x)-
\psi^{+}_{L\sigma}(x)\partial_{x}\psi_{L\sigma}(x))
\label{1}\end{equation}
\begin{equation}
H_{I}=\frac{1}{2}V\sum_{\sigma}\int dx(\rho_{R\sigma}(x)+\rho_{L\sigma}(x)
)^{2}
\label{2}\end{equation}
\begin{equation}
H_{K}=J_{0}(\hat{s}_{R}(0)+\hat{s}_{L}(0))\cdot\hat{S}+J_{2k_{F}}
(\hat{s}_{RL}(0)+\hat{s}_{LR}(0))\cdot\hat{S}
\label{3}\end{equation}
where $\hat{s}_{R(L)}(x)=\frac{1}{2}\psi^{+}_{R(L)\alpha}(x)\hat{\sigma}
_{\alpha\beta}\psi_{R(L)\beta}(x)$, $\hat{s}_{RL}(x)=\frac{1}{2}\psi^{+}
_{R\alpha}(x)\hat{\sigma}_{\alpha\beta}\psi_{L\beta}(x)$,\\ $\hat{s}_{LR}(x)=
\frac{1}{2}\psi^{+}_{L\alpha}(x)\hat{\sigma}_{\alpha\beta}\psi_{R\beta}(x)$; 
$\psi_{R\sigma}(x)$ are the field operator of the electrons that propagate to
the right with wave vectors $\sim +k_{F}$; $\psi_{L\sigma}(x)$ are the field
operators of left propagating electrons with wave vectors $\sim -k_{F}$; $V$
describes density-density interaction with same spin direction with momentum
transferring much less than $k_{F}$. It will be assumed hereafter that the
position of the magnetic impurity is fixed at $x=0$; $J_{0}$ and $J_{2k_{F}}$
are the forward and backward
scattering potential, respectively.
Here, for simplicity we only consider the electron-electron interactions such
as that of Eq.(\ref{2}). The Hamiltonian $H_{0}+H_{I}$ can be derived from a 
lattice model:
$H=t\sum_{<ij>}\sum_{\sigma}[c^{+}_{i\sigma}c_{j\sigma}+h.c.]+
V_{0}\sum_{ij}\sum_{\sigma}n_{i\sigma}n_{j\sigma}$, where
$n_{i\sigma}=c^{+}_{i\sigma}c_{i\sigma}$, by decomposing the
electron operator into as $c_{i\sigma}=e^{ik_{F}x_{i}}\psi_{R\sigma}(x_{i})
+e^{-ik_{F}x_{i}}\psi_{L\sigma}(x_{i})$, and neglecting the Umklapp term 
(assuming
far away from half filling). Of course this model is less popular than the 
usual Hubbard model, i.e., by considering the interaction: $\bar{V}\int dx
(\rho_{R\uparrow}(x)+\rho_{L\uparrow}(x))(\rho_{R\downarrow}(x)+
\rho_{L\downarrow}(x))$. The only difference between them is that in the 
former case the interactions in the charge and 
spin channels are repulsive; in the latter case the interaction in the
charge channel is repulsive, while in the spin channel it is
attractive. This difference significantly influences the behavior of the
impurity susceptibility (see below).  
For a general 1D electronic system, we should consider the interaction:
$\sum_{\sigma}\int dx\{V_{1}\rho_{R\sigma}(x)\rho_{L\sigma}(x)
+\bar{V}_{1}\rho_{R\sigma}(x)\rho_{L-\sigma}(x)
+V_{2}[\rho^{2}_{R\sigma}+\rho^{2}_{L\sigma}]
+\bar{V}_{2}[\rho_{R\sigma}\rho_{R-\sigma}+
\rho_{L\sigma}\rho_{L-\sigma}]\}$. If we take $\bar{V}_{1}=\bar{V}_{2}=0$, 
and $V_{2}=V_{1}/2$, it
reduces to that in (3). $\bar{V}_{1}\neq 0$ and $\bar{V}_{2}\neq 0$ correspond
to that the interactions in spin and charge channels are different which can
be described by the parameters $g_{s}$ and $g_{c}$, respectively. Using the
bosonic representation of the electron fields $\psi_{R(L)\sigma}$, we can
easily demonstrate the term $\hat{s}_{RL}(0)+\hat{s}_{LR}(0)$ has a conformal
dimension $g$ (or generally, $(g_{c}+g_{s})/2$). For the repulsive
electron-electron interaction, the last term in (4) is relevant, the backward
scattering potential is therefore renormalized to be infinity.  

To more effectively treat the backward scattering term, 
we define a set of new fermion operators
\begin{eqnarray}
\psi_{1\sigma}(x) &= & \displaystyle{\frac{1}{\sqrt{2}}(\psi_{R\sigma}(x)+
\psi_{L\sigma}(-x))}  \nonumber \\
\psi_{2\sigma}(x) &= & \displaystyle{\frac{1}{\sqrt{2}}(\psi_{R\sigma}(x)-
\psi_{L\sigma}(-x))}
\label{4}\end{eqnarray}
It is easy to check that the operators $\psi_{1(2)\sigma}(x)$ have the
standard anticommutation relations. In terms of 
these new fermion operators, the
Hamiltonians (\ref{1}) and (\ref{3}) can be rewritten as 
\begin{equation}
H_{0}=-iv_{F}\int dx(\psi^{+}_{1\sigma}(x)\partial_{x}\psi_{1\sigma}(x)+
\psi^{+}_{2\sigma}(x)\partial_{x}\psi_{2\sigma}(x))
\label{5}\end{equation}
\begin{equation}
H_{K}=J_{0}(\hat{s}_{1}(0)+\hat{s}_{2}(0))\cdot\hat{S}+J_{2k_{F}}(
\hat{s}_{1}(0)-\hat{s}_{2}(0))\cdot\hat{S}
\label{6}\end{equation}
where $\hat{s}_{1(2)}(x)=\frac{1}{2}\psi^{+}_{1(2)\alpha}\hat{\sigma}_{\alpha
\beta}\psi_{1(2)\beta}(x)$.
It is worth notice that for a free electron gas, the system reduces to an
anisotropic two-channel Kondo model, the backward scattering potential
$J_{2k_{F}}$ produces the channel anisotropy which completely destroys the
non-Fermi liquid behavior of the isotropic two-channel Kondo model\cite{b19}.
The bosonic representation of the fermion operators
$\psi_{1(2)\sigma}(x)$ can be written in the standard bosonization
technique[\cite{23}-\cite{25}]
\begin{equation}
\psi_{1(2)\sigma}(x)=\frac{1}{\sqrt{2\pi\eta}}\exp\{\frac{2\pi}{L}
\sum_{p\neq 0}\frac{e^{-\frac{\eta}{2}|p|-ipx}}{p}\rho_{1(2)\sigma}(p)\}
=\frac{1}{\sqrt{2\pi\eta}}e^{-i\Phi_{1(2)\sigma}(x)}
\label{6'}\end{equation}
where $\eta$ is an ultraviolet cutoff, $L$ is the length of the system,
$\rho_{1(2)\sigma}(x)=\psi^{+}_{1(2)\sigma}(x)\psi_{1(2)\sigma}(x)$ are the
density operators of the fermion operators $\psi_{1(2)\sigma}(x)$ that have
the same commutation relations as that of the density operators of the
right-branch electrons. 

According to Eq.(\ref{4}), the Hamiltonian (\ref{2}) can be written as
\begin{eqnarray}
H_{I} &=& \displaystyle{ \frac{V}{4}\sum_{\sigma}\int dx\{
[\rho_{1\sigma}(x)+\rho_{2\sigma}(x)]^{2}+
[\rho_{1\sigma}(x)+\rho_{2\sigma}(x)]
[\rho_{1\sigma}(-x)+\rho_{2\sigma}(-x)]} \label{2'} \\
&+& \displaystyle{ [\psi^{+}_{1\sigma}(x)\psi_{2\sigma}(x)+
\psi^{+}_{2\sigma}(x)\psi_{1\sigma}(x)]^{2}} \nonumber \\
&-& \displaystyle{
[\psi^{+}_{1\sigma}(x)\psi_{2\sigma}(x)+
\psi^{+}_{2\sigma}(x)\psi_{1\sigma}(x)]
[\psi^{+}_{1\sigma}(-x)\psi_{2\sigma}(-x)+
\psi^{+}_{2\sigma}(-x)\psi_{1\sigma}(-x)]} 
\nonumber\end{eqnarray}
Using the bosonization representations of the fermion operators
$\psi_{1(2)\sigma}(x)$, the Hamiltonians (\ref{2'}) and (\ref{5}) can be
written as 
\begin{eqnarray}
H &= & \displaystyle{H_{0}+H_{I}}  \nonumber \\
&= & \displaystyle{
\frac{v_{F}}{4\pi(1-\gamma)}\int dx[(\Phi^{'}_{+c}(x))^{2}+(\Phi^{'}_{+s}(x)
)^{2}+\gamma\Phi^{'}_{+c}(x)\Phi^{'}_{+c}(-x)+\gamma\Phi^{'}_{+s}(x)
\Phi^{'}_{+s}(-x)]}  \nonumber \\
&+ & \displaystyle{
\frac{v_{F}}{4\pi}\int dx[(\Phi^{'}_{-c}(x))^{2}+(\Phi^{'}_{-s}(x))^{2}]}
\label{13} \\
&+ & \displaystyle{ \frac{V}{4}\sum_{\sigma}\int dx \{[\psi^{+}_{1\sigma}(x)
\psi_{2\sigma}(x)+\psi^{+}_{2\sigma}(x)\psi_{1\sigma}(x)]^{2}} \nonumber \\
&-& \displaystyle{
[\psi^{+}_{1\sigma}(x)
\psi_{2\sigma}(x)+\psi^{+}_{2\sigma}(x)\psi_{1\sigma}(x)]
[\psi^{+}_{1\sigma}(-x)
\psi_{2\sigma}(-x)+\psi^{+}_{2\sigma}(-x)\psi_{1\sigma}(-x)]\}}
\nonumber\end{eqnarray}
where $\gamma=\frac{V}{2\pi v_{F}+V}$, $\Phi^{'}_{\pm c(s)}(x)=\frac{\partial
\Phi_{\pm c(s)}(x)}{\partial x}=2\pi\rho_{\pm c(s)}(x)$, 
$\Phi_{\pm c}(x)=\frac{1}{2}[\Phi_{\pm\uparrow}(x)+\Phi_{\pm\downarrow}(x)]$,
$\Phi_{\pm s}(x)=\frac{1}{2}[\Phi_{\pm\uparrow}(x)-\Phi_{\pm\downarrow}(x)]$,
$\Phi_{\pm\sigma}(x)=\Phi_{1\sigma}(x)\pm\Phi_{2\sigma}(x)$
$\rho_{\pm c}(x)=
\frac{1}{2}[\rho_{\pm\uparrow}(x)+\rho_{\pm\downarrow}(x)]$, $\rho_{\pm s}(x)=
\frac{1}{2}[\rho_{\pm\uparrow}(x)-\rho_{\pm\downarrow}(x)]$, $\rho_{\pm\sigma}
(x)=\rho_{1\sigma}(x)\pm\rho_{2\sigma}(x)$, 
The fields $\Phi_{+c(s)}(x)$ reduce into free boson fields, while the fields
$\Phi_{-c(s)}(x)$ are highly self-interacting boson fields because of the last
two terms. For simplicity, we have not apparently written out the boson fields
$\Phi_{-c(s)}(x)$ in the last terms which are not needed in the following
calculation of the impurity susceptibility. However, it is easily to show that
the last two terms are
independent of the boson fields $\Phi_{+c(s)}(x)$. If we define two
new parameters
\begin{equation}
J^{\bot}_{0}=J_{1}+J_{2},\;\;\; J^{\bot}_{2k_{F}}=J_{1}-J_{2}
\label{10}\end{equation}
the Hamiltonian (\ref{6}) can be written in the bosonization representation as
\begin{eqnarray}
H_{K} &= & \displaystyle{
\frac{2\delta_{+}v_{F}}{\pi}\Phi^{'}_{+s}(0)S^{z}+\frac{2\delta_{-}v_{F}}{\pi}
\Phi^{'}_{-s}(0)S^{z}}  \nonumber \\
&+ & \displaystyle{
\frac{J_{1}}{2\pi\eta}[e^{-i\Phi_{+s}(0)}e^{-i\Phi_{-s}(0)}S^{+}+
e^{i\Phi_{+s}(0)}e^{i\Phi_{-s}(0)}S^{-}]} \nonumber  \\
&+ & \displaystyle{
\frac{J_{2}}{2\pi\eta}[e^{-i\Phi_{+s}(0)}e^{i\Phi_{-s}(0)}S^{+}
+e^{i\Phi_{+s}(0)}e^{-i\Phi_{-s}(0)}S^{-}]}
\label{13'}\end{eqnarray}
where $\delta_{+}=\arctan(\frac{J^{z}_{0}}{4v_{F}})$, $\delta_{-}=\arctan(
\frac{J^{z}_{2k_{F}}}{4v_{F}})$. These definitions of the phase shifts
$\delta_{+}$ and $\delta_{-}$ stem from the exact solution of X-ray
absorption problem\cite{26}, they are valid both for small and large 
$J^{z}_{0}$ and $J^{z}_{2k_{F}}$. It is very clear that only the boson fields
$\Phi_{\pm s}(x)$ of the spin part interact with the impurity spin, therefore,
the electron-electron interaction existing in the spin channels can influence
the low temperature behavior of the impurity susceptibility, while the 
electron-electron
interaction existing in the charge channels cannot influence the low
temperature behavior of the impurity susceptibility in the Toulouse limit, 
such as the 
density-density interaction:
$V\sum_{\sigma\sigma^{'}}\int dx (\rho_{R\sigma}(x)+\rho_{L\sigma}(x))
(\rho_{R\sigma^{'}}(x)+\rho_{L\sigma^{'}}(x))$, we can easily prove that this
type electron-electron interaction only induces the interactions in the charge
channels described by the boson fields $\Phi_{\pm c}(x)$. 
It is worth notice
that the spin and charge channels we used are described by the boson
fields $\Phi_{\pm s}(x)$ and $\Phi_{\pm c}(x)$, respectively, that may have a
little difference from the real electron spin and charge channels due to the
definitions of the fermion operators in (\ref{4}). 

For simplicity, first we consider a free electron gas, $\gamma=0$, the
Hamiltonian (\ref{13}) reduces into as
\begin{equation}
H_{0}=\frac{v_{F}}{4\pi}\int dx [(\Phi^{'}_{+c}(x))^{2}+
(\Phi^{'}_{+s}(x))^{2}+(\Phi^{'}_{-c}(x))^{2}+(\Phi^{'}_{-s}(x))^{2}]
\label{14}\end{equation}
If we take the following unitary transformation
\begin{equation}
U=\exp\{i\frac{2\delta_{+}}{\pi}\Phi_{+s}(0)S^{z}+i\frac{2\delta_{-}}{\pi}
\Phi_{-s}(0)S^{z}\}
\label{15}\end{equation}
we can cancel the $\delta_{+}$ and $\delta_{-}$ terms in (\ref{13'}), and the
total Hamiltonian $H_{0}+H_{K}$ can be written as
\begin{eqnarray}
H^{'} &= & \displaystyle{U^{+}(H_{0}+H_{K})U}  \nonumber \\
&= & \displaystyle{
H_{0}+\frac{J_{1}}{2\pi\eta}}[
\displaystyle{\exp\{-i(\frac{2\delta_{+}}{\pi}+1)\Phi_{+s}(0)
-i(\frac{2\delta_{-}}{\pi}+1)\Phi_{-s}(0)\}\cdot S^{+}} \nonumber  \\
&+ & \displaystyle{
\exp\{i(\frac{2\delta_{+}}{\pi}+1)\Phi_{+s}(0)+i(\frac{2\delta_{-}}{\pi}+1)
\Phi_{-s}(0)\}\cdot S^{-}}]  \nonumber \\
&+ & \displaystyle{\frac{J_{2}}{2\pi\eta}}[\displaystyle{
\exp\{-i(\frac{2\delta_{+}}{\pi}+1)\Phi_{+s}(0)+i(1-\frac{2\delta_{-}}{\pi})
\Phi_{-s}(0)\}\cdot S^{+}}  \nonumber \\
&+ & \displaystyle{
\exp\{i(\frac{2\delta_{+}}{\pi}+1)\Phi_{+s}(0)-i(1-\frac{2\delta_{-}}{\pi})
\Phi_{-s}(0)\}\cdot S^{-}}]
\label{16}\end{eqnarray}
In the strong coupling limit (Toulouse limit), the phase shifts can take
the following values
\begin{equation}
\delta^{c}_{+}=-\frac{\pi}{2}, \;\;\; \delta^{c}_{-}=\pm\frac{\pi}{2}
\label{17}\end{equation}
For the case of $\delta^{c}_{+}=-\frac{\pi}{2}$, 
$\delta^{c}_{-}=-\frac{\pi}{2}$, the total
Hamiltonian (\ref{16}) reduces into as
\begin{eqnarray}
H^{'}_{c} &= & \displaystyle{H_{0}+\frac{J_{1}}{2\pi\eta}(S^{+}+S^{-})}
\nonumber \\
&+ & \displaystyle{\frac{J_{2}}{2\pi\eta}[e^{i2\Phi_{-s}(0)}S^{+}+
e^{-i2\Phi_{-s}(0)}S^{-}]}
\label{18}\end{eqnarray}
For the case of $\delta^{c}_{+}=-\frac{\pi}{2}$, 
$\delta^{c}_{-}=\frac{\pi}{2}$, the total
Hamiltonian (\ref{16}) can be written as
\begin{eqnarray}
H^{'}_{c} &= & \displaystyle{H_{0}+\frac{J_{2}}{2\pi\eta}(S^{+}+S^{-})}
\nonumber \\
&+ & \displaystyle{\frac{J_{1}}{2\pi\eta}[e^{-i2\Phi_{-s}(0)}S^{+}+
e^{i2\Phi_{-s}(0)}S^{-}]}
\label{19}\end{eqnarray}
It is worth notice that the violation of the $SU(2)$ symmetry of the system in
the strong coupling limit is artificial because in the bosonization
representation of the fermion fields (8) we have omitted the constant fermion
operators $\hat{U}_{1(2)\sigma}$ which guarantee the anticommutation relation
of the fermion fields $\psi_{1(2)\sigma}(x)$. In fact, we still have the
$SU(2)$ symmetry in the strong coupling limit.
If we define the impurity spin $\hat{S}$ as: $S^{+}=f^{+}$, $S^{-}=f$, $S^{z}=
f^{+}f-1/2$, the Hamiltonians (\ref{18}) and (\ref{19}) are very similar to
that in Ref.\cite{m} derived from the quantum dot.
The $J_{1}$ (the former case) or $J_{2}$
(the latter case) term provides an energy gap $\Delta \sim J_{1}\;(or\; 
J_{2})$ to
the impurity fermion $f$. It means that in the strong coupling
limit, the impurity fermion $f$ form a spin singlet (Kondo singlet)
with the conduction electrons at the impurity site $x=0$. Therefore the system
has the usual Fermi liquid behavior. This property of the system is very
simple and clear, 
because for a free electron gas, the system becomes an usual anisotropic
two-channel Kondo model, the backward scattering potential $J_{2k_{F}}$
produces the channel anisotropy which destroys the non-Fermi liquid behavior
of the isotropic two-channel Kondo model and makes the system show the 
usual Fermi
liquid behavior. It is reasonable for choosing the phase shift values in
(\ref{17}) in the strong coupling limit, for example, for an isotropic case
(i.e., $J_{2k_{F}}=0$) we have the relations: $\delta^{c}_{+}=-\frac{\pi}{2}$,
$\delta_{-}\equiv 0$, $ J_{1}\equiv J_{2}$, the Hamiltonian (\ref{16}) becomes
the well-known Hamiltonian derived by Emery and Kivelson\cite{27} from the
isotropic two-channel Kondo model.
On the other hand, for the most anisotropic case (i.e., $J_{0}=0$) we have the
relations: $\delta_{+}\equiv 0$, $\delta_{-}=\pm\frac{\pi}{2}$, $J_{2}\equiv
-J_{1}$, the Hamiltonian (\ref{16}) becomes the well-known resonant-level
model induced by the one-channel Kondo model in the Toulouse limit. The
definition of the Kondo interaction in Eq.(3) means that $J^{z}_{0}>0$
corresponds to antiferromagnetic exchange, in the Toulouse limit the phase
shift $\delta_{+}$ would take the value $+\pi/2$, why do we take it as
$-\pi/2$? The reason is that first in the case of $J_{2k_{F}}$,
$\delta^{c}_{+}=-\pi/2$ reproduces the famous form of the isotropic
two-channel Kondo model obtained by Emery and Kivelson. Second, in the
bosonization description of the Kondo interaction (12), it is the way to
incorporate the antiferromagnetic exchange by taking $J^{z}_{0}<0$ because the
first term in (12) becomes simple potential scattering and only describes the
same direction spin-spin interaction.

For an interacting electron gas, we can take the following unitary
transformation
\begin{equation}
U=\exp\{i\frac{2\delta_{+}}{\pi}g^{2}\Phi_{+s}(0)S^{z}+
i\frac{2\delta_{-}}{\pi}\Phi_{-s}(0)S^{z}\}
\label{20}\end{equation}
to eliminate the $\delta_{+}$ and $\delta_{-}$ terms in (\ref{13'}), where
$g=(\frac{1-\gamma}{1+\gamma})^{1/2}$ is a dimensionless coupling strength
parameter.    
Under this unitary transformation (\ref{20}), the $J_{1}$ and $J_{2}$ terms in
(\ref{13'}) can be written as
\begin{eqnarray}
\displaystyle{\frac{J_{1}}{2\pi\eta}} && [\displaystyle{
\exp\{-i(\frac{2\delta_{+}}{\pi}g^{2}+1)\Phi_{+s}(0)-i
(\frac{2\delta_{-}}{\pi}+1)\Phi_{-s}(0)\}\cdot S^{+}}  \nonumber \\
+ && \displaystyle{
\exp\{i(\frac{2\delta_{+}}{\pi}g^{2}+1)\Phi_{+s}(0)+i
(\frac{2\delta_{-}}{\pi}+1)\Phi_{-s}(0)\}\cdot S^{-}}]  \nonumber \\
+ && \displaystyle{\frac{J_{2}}{2\pi\eta}}[\displaystyle{
\exp\{-i(\frac{2\delta_{+}}{\pi}g^{2}+1)\Phi_{+s}(0)+i
(1-\frac{2\delta_{-}}{\pi})\Phi_{-s}(0)\}\cdot S^{+}}  \nonumber \\
+ && \displaystyle{
\exp\{i(\frac{2\delta_{+}}{\pi}g^{2}+1)\Phi_{+s}(0)-i
(1-\frac{2\delta_{-}}{\pi})\Phi_{-s}(0)\}\cdot S^{-}}]
\label{21}\end{eqnarray}
If we take the following gauge transformations: 
\begin{equation}
\psi_{1(2)\sigma}(x)=\bar{\psi}_{1(2)\sigma}(x)e^{i\sigma\theta_{1(2)}}
,\;\;\; \theta_{1}-\theta_{2}=2\delta_{-}S^{z}
\label{gauge}\end{equation}
in the strong coupling critical point defined by the backward scattering
potential $J_{2k_{F}}$: $\delta^{c}_{-}=\pm\pi/2$, 
the Hamiltonian (\ref{13}) can be written as
\begin{eqnarray}
H &= & \displaystyle{
\frac{v_{F}}{4\pi(1-\gamma)}
\int dx[(\bar{\Phi}^{'}_{+c}(x))^{2}+(\bar{\Phi}^{'}
_{+s}(x))^{2}+\gamma\bar{\Phi}^{'}_{+c}(x)\bar{\Phi}^{'}_{+c}(-x)+\gamma
\bar{\Phi}^{'}_{+s}(x)\bar{\Phi}^{'}_{+s}(-x)]}  \nonumber \\
&+ & \displaystyle{
\frac{v_{F}}{4\pi}\int dx[(\bar{\Phi}^{'}_{-c}(x))^{2}+(\bar{\Phi}^{'}
_{-s}(x))^{2}]} \nonumber  \\
&+ & \displaystyle{ \frac{V}{4}\sum_{\sigma}\int dx\{
[\bar{\psi}^{+}_{1\sigma}(x)\bar{\psi}_{2\sigma}(x)+
\bar{\psi}^{+}_{2\sigma}(x)\bar{\psi}_{1\sigma}(x)]^{2}} \label{88} \\
&+ & \displaystyle{
[\bar{\psi}^{+}_{1\sigma}(x)\bar{\psi}_{2\sigma}(x)+
\bar{\psi}^{+}_{2\sigma}(x)\bar{\psi}_{1\sigma}(x)]
[\bar{\psi}^{+}_{1\sigma}(-x)\bar{\psi}_{2\sigma}(-x)+
\bar{\psi}^{+}_{2\sigma}(-x)\bar{\psi}_{1\sigma}(-x)]\}} \nonumber
\end{eqnarray}
where, $\bar{\Phi}_{\pm
c}(x)=\frac{1}{2}[\bar{\Phi}_{\pm\uparrow}(x)+\bar{\Phi}_{\pm
\downarrow}(x)]$, $\bar{\Phi}_{\pm
s}(x)=\frac{1}{2}[\bar{\Phi}_{\pm\uparrow}(x)-\bar{\Phi}_{\pm
\downarrow}(x)]$,\\
$\bar{\Phi}_{\pm\sigma}(x)=\bar{\Phi}_{1\sigma}(x)\pm\bar{\Phi}_{2
\sigma}(x)$; the
bosonic representation of the fermion fields $\bar{\psi}_{1(2)\sigma}(x)$
are $\bar{\psi}_{1(2)\sigma}(x)=
(\frac{1}{2\pi\eta})^{1/2}\exp\{-i\bar{\Phi}_{1(2)\sigma}(x)\}$.
The critical points $\delta^{c}_{-}=\pm\pi/2$ can be reached for $g\leq 1$ (or
generally, $(g_{c}+g_{s})/2\leq 1$) because the backward scattering potential
is renormalized to be infinity in the low energy limit.
It is worth notice that the last term in (\ref{88}) changes sign after
performing the unitary and gauge transformations and taking the strong
coupling limit of the backward scattering potential.  

For an attractive electron-electron interaction in the spin channels, 
$g_{s}>1$, but keeping $(g_{c}+g_{s})/2\leq 1$, in the strong coupling
limit, the phase shifts can take following values
\begin{equation}
\delta^{c}_{+}=-\frac{\pi}{2g^{2}_{s}}, \;\;\; \delta^{c}_{-}=\pm\frac{\pi}{2}
\label{22}\end{equation}
$(g_{c}+g_{s})/2\leq 1$ guaratees the last equation to be valid, and the
attractive interaction in the spin channels enhances the effective "scattering
channels" and the phase shift $\delta_{+}$ takes less value than $\pi/2$.
In this case, the impurity part shows the same low energy behavior as that for
the free electron system. However, with the interaction in Eq.(3),
$g_{c}=g_{s}=g>1$, we cannot reach the critical points
$\delta^{c}_{-}=\pm\frac{\pi}{2}$, because in this case the backward
scattering potential is renormalized to be zero in the low energy limit. 

For a repulsive
electron-electron interaction, $g<1$, in the strong coupling limit, the phase
shifts can only take the following values
\begin{equation}
\delta^{c}_{+}=-\frac{\pi}{2}, \;\;\; \delta^{c}_{-}=\pm\frac{\pi}{2}
\label{23}\end{equation}
therefore the $J_{1}$ or $J_{2}$ term in (\ref{21}) can be written as
\begin{equation}
\frac{J}{2\pi\eta}[e^{i(g^{2}-1)\Phi_{+s}(0)}S^{+}+e^{-i(g^{2}-1)\Phi_{+s}(0)}
S^{-}]
\label{24}\end{equation}
where $J=J_{1}$ for the case of $\delta^{c}_{+}=\frac{\pi}{2}, \; 
\delta^{c}_{-}=\frac{\pi}{2}$;
$J=J_{2}$ for the case of $\delta^{c}_{+}=\frac{\pi}{2}, \;
\delta^{c}_{-}=-\frac{\pi}{2}$. We have
omitted the high order terms. It is worth notice that the gauge
transformations (\ref{gauge}) retain the boson field $\Phi_{+s}(x)$
invariance, i.e., $\Phi_{+s}(x)=\bar{\Phi}_{+s}(x)$. 
If we define an anyon field:
$\psi(x)=\frac{1}{\sqrt{2\pi\eta}}e^{-i(g^{2}-1)\bar{\Phi}_{+s}(x)}$ 
and impurity
fermion operators: $S^{+}=f^{+}$, $S^{-}=f$, $S^{z}=f^{+}f-\frac{1}{2}$, then
the equation (\ref{24}) can be rewritten as
\begin{equation}
\frac{J}{\sqrt{2\pi\eta}}[f^{+}\psi(0)+\psi^{+}(0)f]
\label{25}\end{equation}
According to Eq.(\ref{88}), we can easily obtain following
correlation functions
\begin{eqnarray}
\displaystyle{
<e^{-i\Phi_{+s}(0,t)}e^{i\Phi_{+s}(0,0)}>} && \displaystyle{
\sim(\frac{1}{t})^{\frac{1}{g}}}  \nonumber \\
<\psi(0,t)\psi^{+}(0,0)> && \sim
\displaystyle{(\frac{1}{t})^{\frac{1}{g}(1-g^{2})^{2}}}
\label{26}\end{eqnarray}
However, using Eq.(\ref{25}), we can easily calculate the self-energy
$\Sigma(\omega)$ of the impurity fermion $f$ by the correlation function of
the anyon field $\psi(0,t)$:
$\Sigma(\omega)\sim\omega^{-1+(1-g^{2})^{2}/g}$. The Green's function of the
impurity fermion $f$ is $1/G(\omega)=i\omega+\Sigma(\omega)$. Therefore, in
the long time limit (i.e., in the low energy limit), we have the following
asymptotic behavior which significantly depends on the dimensionless coupling
strength parameter $g$
\begin{equation}
<f(t)f^{+}(0)> \sim \left\{\begin{array}{ll}
\displaystyle{(\frac{1}{t})^{2-\frac{1}{g}(1-g^{2})^{2}}}, & \;\; g_{c}<g<1\\
\displaystyle{e^{-i\epsilon_{f}t}}, & \;\; g\leq g_{c}\end{array}\right.
\label{27}\end{equation}
where $g_{c}$ is defined as: $(1-g^{2}_{c})^{2}=2g_{c}$, $\epsilon_{f}$ is the
Fermi level of the impurity fermion $f$. It is very clear that the physical
interpretation of this special coupling constant $g_{c}$ is that at this point
the self-energy of the impurity fermion contributed by the conduction electrons
has a linear frequency dependence.
It is very surprising that for a
strong repulsive electron-electron interaction, $g\leq g_{c}$, the impurity
fermion $f$ becomes a free fermion in the low energy and low temperature
limit. Eq. (\ref{27}), the central result of present paper, is clearly shown
that the non-Fermi liquid behavior of the impurity susceptibility in the 
TL-liquid completely stems from the coupling between the 
impurity spin and the total spin freedom
degree of the conduction electrons described by the boson field $\Phi_{+s}(x)$
(total spin channel). It is independent of the coupling existing in the charge
channels of the conduction electrons described by the boson fields $\Phi_{\pm
c}(x)$ because in the representation of the fermions $\psi_{1(2)\sigma}(x)$
the Kondo interaction term $H_{K}$ (\ref{13'}) is not coupling with the boson
fields $\Phi_{\pm c}(x)$. Therefore, for choosing different type
electron-electron interaction, one may obtain a Fermi liquid fixed point or a
non-Fermi liquid fixed point in the strong coupling limit by the perturbation
method such as the renormalization group.
For the case of $g=1$, the impurity spin forms a Kondo singlet with the spin 
freedom
degrees of the conduction electrons described by the boson fields $\Phi_{\pm
s}(x)$, and it is completely screened by the conduction electrons. As a whole,
they show a non-magnetic impurity behavior at the impurity site, therefore,
the system has the usual Fermi liquid behavior. For the case of $g<1$, in the 
flavor-spin
channel the boson field $\Phi_{-s}(x)$ and the impurity spin form a bound
state at the impurity site which induces the impurity spin 
decoupling from the flavor-spin channel
described by the boson field $\Phi_{-s}(x)$ in the strong coupling limit. On
the other hand, in the total spin channel the boson field $\Phi_{+s}(x)$  
and the impurity spin still form a bound state at the impurity site, due
to the repulsive interactions among the electrons there exists a net coupling
between the total spin channel described by the boson field $\Phi_{+s}(x)$
and the impurity spin. 
This unusual behavior of the impurity spin comes from that because of the
repulsive electron-electron interaction the density of state of the total spin
collective mode described by the boson field $\Phi_{+s}(x)$ is decreasing as
the dimensionless coupling strength parameter $g$ is decreasing. Therefore, it
cannot completely screen the impurity spin in the total spin channel. As
$g\leq g_{c}$, in the total spin channel the impurity fermion shows a free
fermion behavior in the low energy and low temperature limit. According to
Eq. (\ref{27}), we can easily obtain the temperature dependence of the
impurity susceptibility 
\begin{equation}
\chi_{im}(T)\sim\left\{\begin{array}{ll}
\displaystyle{T^{3-\frac{2}{g}(1-g^{2})^{2}}}, & \;\;\; g_{c}<g<1\\
\displaystyle{\frac{1}{T}}, & \;\;\; g\leq g_{c}
\end{array}\right.
\label{28}\end{equation}
which shows a power law non-Fermi liquid behavior. It is worth notice that all
above discussions are confined in the strong coupling region determined by the
phase factors $\delta^{c}_{\pm}$. We can determine the low energy behavior of
the impurity by considering the leading irrelevant terms in this strong
coupling region $\Delta H=\lambda \Phi^{'}_{+s}(0)S^{z}+\tilde{\lambda}
\Phi^{'}_{-s}(0)S^{z}$, where $\lambda$ and $\tilde{\lambda}$ are small
coupling constants. It is nontrivial to get the correlation function
of the boson field $\Phi^{'}_{-s}(x)$ at the impurity site $x=0$ due to the
relation
$\Phi^{'}_{-s}(0)/\pi=\psi^{+}_{R\uparrow}(0)\psi_{L\uparrow}(0)   
+\psi^{+}_{L\uparrow}(0)\psi_{R\uparrow}(0)-
\psi^{+}_{R\downarrow}(0)\psi_{L\downarrow}(0)
-\psi^{+}_{L\downarrow}(0)\psi_{R\downarrow}(0)$, which depends on the
interactions in the charge channels. However, in the strong
coupling region determined by the phase factors $\delta^{c}_{\pm}$ (24), 
the low
temperature behavior of the impurity susceptibility and specific heat is
independent of the interactions existing in the charge channels because in the
case of $g=1$ there exists the gap in the excitation spectrum of the
impurity fermion (see Eqs.(\ref{18}) and (\ref{19})), in the low energy limit,
the specific heat and susceptibility of the impurity are exponentially
decreasing. In the case of $g<1$, 
the boson field $\Phi^{'}_{-s}(0)$ has the correlation function
$<\Phi^{'}_{-s}(0,t)\Phi^{'}_{-s}(0,0)>\sim t^{-2/g}$ (or $\sim
t^{-1/g_{c}-1/g_{s}}$ in the general case) in this strong
coupling region. Therefore, the leading irrelevant term
$\lambda\Phi^{'}_{+s}(0)S^{z}$ is dominant because the boson field    
$\Phi^{'}_{+s}(0)$ has the correlation function 
$<\Phi^{'}_{+s}(0,t)\Phi^{'}_{+s}(0,0)>\sim t^{-2}$. However, if the system is
far away from this strong coupling region determined by the phase factors
$\delta^{c}_{\pm}$, for example, it is in the region controlled by the phase
factors $\delta^{c}_{+}$ and $\delta_{-}= 0$, the specific heat of the
impurity can be influenced by the interaction in the charge channels, but the
impurity susceptibility is still independent of the interaction in the charge
channels. 

In summary, by using the bosonization technique, 
we have studied in detail the low
temperature property of the Kondo effect in the TL-liquid and first time shown
that the power law temperature dependence of the impurity susceptibility is
completely determined by the repulsive electron-electron interaction existing
in the total spin channel described by the boson field $\Phi_{+s}(x)$ and is
independent of the electron-electron interaction existing in the charge
channels described by the boson fields $\Phi_{\pm c}(x)$. Therefore, for
choosing different type electron-electron interactions, one may obtain an
usual Fermi liquid fixed point or a non-Fermi liquid fixed point in the
strong coupling limit of the backward scattering potential because they 
completely depend upon the
electron-electron interaction existing in the total spin channel.

The author would like to thank Prof. P. Fulde for encouragement.


\newpage
\begin{references}

\bibitem{1} T.Ogawa, A.Furusaki, and N.Nagaosa, Phys. Rev. Lett. {\bf 68}, 
3638(1992); A.Furusaki, and N.Nagaosa, Phys. Rev. {\bf B}47, 3827(1993).

\bibitem{2} D.K.K.Lee, and Y.Chen, Phys. Rev. Lett. {\bf 69}, 1399(1992).

\bibitem{3} C.L.Kane, and M.P.A.Fisher, Phys. Rev. Lett. {\bf 68}, 
1220(1992); Phys. Rev. {\bf B}46, 15233(1992).

\bibitem{3'} X.G.Wen, Phys. Rev. {\bf B}41, 12838(1990);
Int. J. Mod. Phys. {\bf B}6, 1711(1992).

\bibitem{4} K.A.Matveev, D.X.Yue, and L.I.Glazman, Phys. Rev. Lett. {\bf 
71}, 3351(1993).

\bibitem{5} A.D.Gogolin, Phys. Rev. Lett. {\bf 71}, 2995(1993).

\bibitem{6} N.V.Prokof'ev, Phys. Rev. {\bf B}49, 2148(1994).

\bibitem{7} C.L.Kane, K.A.Matveev, and L.I.Glazman, Phys. Rev. {\bf B}49, 
2253(1994).

\bibitem{8} I.Affleck, and A.W.W.Ludwig, J. Phys. {\bf A}27, 5375(1994).

\bibitem(9) M.Ogata, and H.Fukuyama, Phys. Rev. Lett. {\bf 73}, 468(1994).

\bibitem{10} P.Fendley, A.W.W.Ludwig, and H.Saleur, Phys. Rev. Lett. 
{\bf 74}, 3005(1995).

\bibitem{11} F.Guinea, G.G\'{o}mez-Santos, M.Sassetti, and M.Ueda, 
Europhys. Lett. {\bf 30}, 561(1995).

\bibitem{12} S.Tarucha, T.Honda, and T.Saku, Solid State Commun., {\bf 
94}, 413{1995}.

\bibitem{13} K.Moon {\it et al}., Phys. Rev. Lett. {\bf 71}, 4381(1993); 
K.Moon, and S.M.Girvin, cond-mat/9511013.

\bibitem{13'} F.P.Milliken, C.P.Umbach, and R.A.Webb, Solid State Commun., 
{\bf 97}, 309(1996).

\bibitem{14} Y.Oreg, and A.M.Finkel'stein, Phys. Rev. {\bf B}53, 10928(1996);
Phys. Rev. Lett. {\bf 76}, 4230(1996).

\bibitem{15} F.Lesage, H.Saleur, and S.Skorik, Phys. Rev. Lett. {\bf 76}, 
3388(1996); G.G\'{o}mez-Santos, {\it ibid}, {\bf 76}, 4223(1996).

\bibitem{16} Y.L.Liu, Quantum impurity scattering of Tomonaga-Luttinger
liquid, preprint.

\bibitem{b19} M.Fabrizio, A.O.Gogolin, and P.Nozi\`{e}res,
Phys. Rev. Lett. {\bf 74}, 4503(1995);\\
N.Andrei, and A.Jerez, {\it ibid.}, {\bf 74}, 4507(1995).

\bibitem{19} D.H.Lee, and J.Toner, Phys. Rev. Lett. {\bf 69}, 3378(1992).

\bibitem{20} A.Furusaki, and N.Nagaosa, Phys. Rev. Lett. {\bf 72}, 892(1994).

\bibitem{21} A.Schiller, and K.Ingersent, Phys. Rev. {\bf B}51, 4676(1995).

\bibitem{22} P.Fr\"{o}jdh, and H.Johannesson, Phys. Rev. Lett. {\bf 75},
300(1995). 

\bibitem{23} A.Luther, and I.Peschel, Phys. Rev. {\bf B}9, 2911(1974);
{\bf B}12, 3908(1975).

\bibitem{24} V.J.Emery, in {\it Highly Conducting One-Dimensional Solids},
Edited by J.T.Devreese {\it et al}., (Plenum, New York, 1979); J.S\'{o}lyom, 
Adv. Phys. {\bf 28}, 201(1979).

\bibitem{25} F.D.M.Haldane, J. Phys. {\bf C}14, 2585(1981).

\bibitem{26} P.Nozi\`{e}res, C. De Dominicis, Phys. Rev. {\bf 178}, 1097(1969).

\bibitem{m} K.A.Matveev, Phys. Rev. {\bf B}51, 1743(1995).

\bibitem{27} V.J.Emery, and S.Kivelson, Phys. Rev. {\bf B}46, 10812(1992).

\end{references}
\end{document}